# Passively Q-switched diode-pumped $Cr^{4+}$:YAG/$Nd^{3+}$:GdVO$_4$ monolithic microchip laser

**Sebastien Forget, Frederic Druon, François Balembois, Patrick Georges**

Laboratoire Charles Fabry de l'Institut d'Optique, UMR 8501 du CNRS, Université Paris-Sud – 91403 Orsay, France

**Nicolas Landru, Jean-Philippe Fève**

JDS Uniphase, 31 Chemin du vieux chêne, 38941 Meylan, France

**Jiali Lin, Zhiming Weng**

Fujian JDSU CASIX, Inc. Fuzhou P.O.Box 1103

Fuzhou Fujian 350014, China

**Abstract** : the realization of high repetition rate passively Q-switched monolithic microlaser is a challenge since a decade. To achieve this goal, we report here on the first passively Q-switched diode-pumped microchip laser based on the association of a Nd:GdVO$_4$ crystal and a $Cr^{4+}$:YAG saturable absorber. The monolithic design consists of 1 mm long 1% doped Nd:GdVO$_4$ optically contacted on a 0.4 mm long $Cr^{4+}$:YAG leading to a plano-plano cavity. A repetition rate as high as 85 kHz is achieved. The average output power is approximately 400 mW for 2.2 W of absorbed pump power and the pulse length is 1.1 ns.

Pulsed solid-state lasers are widely used in scientific, medical, industrial and military systems. For a large number of applications, the requirements in terms of compactness, reliability, efficiency and cost are very important. Consequently, the interest on passively Q-switched diode-pumped solid-state neodymium doped microchip lasers continuously increases since its first realization by Zayhowski et al.[1] ten years ago. Indeed, since the development of reliable solid-state saturable absorbers, passive Q-switching allows the realization of low cost laser sources (avoiding the use of electro- or acousto-optics elements and corresponding electronics and high voltage) and even more compact packaging. Several solid-state saturable absorbers [2] have been developed to replace the dyes used in the past: the most used saturable absorber is undoubtedly $Cr^{4+}$:YAG. This material presents a good ratio between its ground state and excited state cross section [3]. Moreover, this material is optically well known and benefits from excellent optical quality and cost-effective mass production. The recent developments in the field deal essentially with shorter pulses and higher repetition rates as it is required for applications like high resolution ranging and imaging [4]. Pulses as short 148 ps have been demonstrated with $Cr^{4+}$:YAG passively Q-switched Nd:YAG microlaser [5]. To obtain shorter pulses with high repetition rates, the use of Semiconductor Saturable Absorber Mirrors (SESAM) by Keller et al. was a big step forward [6], leading to pulses as short as 37 ps with a repetition rate of 160 khz [7]. The MHz level can even be achieved with longer pulses (but still sub-nanosecond) [8]. However, passively Q-switched microchip lasers using SESAM are not as mature as $Cr^{4+}$:YAG/Nd:YAG microchip lasers (commercially available since many years) and the pulse energy obtained remains relatively low, mainly due to some damage problems on the semiconductor. Moreover, the realization of fully monolithic microchip with SESAMs has not been demonstrated up to now, to our best knowledge.

The highest repetition rate (in stable and single-longitudinal mode operation) achieved with passively Q-switched $Cr^{4+}$:YAG/Nd:YAG microchip lasers remains around 50 kHz [9] under diode pumping, with a pulse duration of 500 ps. There is consequently a challenge for obtaining short pulses and high repetition rate in a microchip configuration and with stable single-longitudinal mode operation. In this paper, we demonstrate for the first time a high repetition rate passively Q-switched monolithic microlaser using a Nd-doped vanadate crystal ($Nd:GdVO_4$).

## 1. Theoretical considerations and state of the art

The repetition rate in a passively Q-switched microchip laser can be simply derived from the rate equations. When the pumping is switched on, the excited state population starts to increase from zero up to a value corresponding to the oscillation threshold. Assuming that the pulse duration is negligible compared to time interval between two consecutive pulses, the pulse period is equal to the time taken by the gain to reach the losses. So, assuming a four level gain medium with no saturation of absorption and a longitudinal pumping on one side, the repetition rate can be written, after a few basic mathematical manipulations, as :

$$f = \frac{1}{-\tau . \ln\left(1 + \frac{\ln(T^2 R_1 R_2)}{2\sigma L . \eta . \tau . n_t \sigma_p I_p}\right)} \quad (1)$$

where $\tau$ is the fluorescence lifetime, $\sigma$ the emission cross section, L the length of the cavity, $\eta$ a factor of overlapping between the pump beam and the cavity beam (taking also the exponential absorption of the pump into account), $n_t$ the total population density, $\sigma_p$ the absorption cross section, $I_p$ the incident pump intensity (in number of photons per second and per surface unit), T the transmission of the saturable absorber (unsaturated) and $R_1$, $R_2$ the reflection coefficients of the two mirror cavity. The cavity parameters (T, $R_1$, $R_2$ and L) are generally chosen to achieve short pulses (one nanosecond or less) leading to short cavity with relatively low transmission of the saturable absorber and low reflection coefficient for the output coupler. The equation (1) shows that some other parameters could be optimized to increase the repetition rate without modification of the pulse duration. Increasing the pump intensity or the overlapping factor η is the first possibility, leading to the choice of a high brightness laser diode as pump source. The second possibility is to choose a gain medium with appropriate parameters. High repetition rate could be achieved with a high doping level ($n_t$) but it would lead to the increase of thermal problems. Using equation (1), it can be shown easily that the lifetime is not a relevant parameter (there is only very little variation of the repetition rate versus the lifetime). The last possibility concerns the choice of high cross sections (emission and absorption) materials. In the case of Nd-doped vanandate crystals, the emission cross section is 3 to 6 times higher than in Nd:YAG and the absorption cross section up to 7 times higher (depending on the matrix $GdVO_4$ or $YVO_4$ and on the crystal orientation) [10].

Moreover, Nd:doped vanadate absorption band matches the emission band of GaAlAs laser diodes and is 2 times broader than the one of Nd:YAG leading to larger effective absorption coefficient (laser diode spectrum can be a few nanometers wide). Then, the replacement of Nd:YAG by Nd:vanadate crystals (Nd:$YVO_4$ or Nd:$GdVO_4$) is a very interesting alternative solution.

However, as emphasized by several authors [11-13], it is crucial to match the so-called "Q-switching criterion" in order to obtain giant pulses: if the gain cross section of the laser crystal is larger than the absorption cross section of the saturable absorber, the energy stored in the cavity might be too small to bleach the saturable absorber, leading to quasi-CW operation or at least broadening of the pulse duration.

Complete theoretical explanation based on the second threshold condition [12] is given by Zheng et al. [14], leading to the following Q-switching criterion:

$$\frac{\alpha_{sa}\sigma_{sa}L_{sa}A_l}{\alpha_l\sigma L_l A_{sa}} > \frac{\gamma(1-\chi)}{1-\left(\frac{\sigma_e}{\sigma_g}\right)_{as}} \quad (2)$$

where $\alpha_{sa}$, $L_{sa}$, $\sigma_{sa}$ and $A_{sa}$ are respectively the absorption coefficient, the length, the absorption cross-section and the effective beam area of the saturable absorber. $\alpha_l$, $L_l$, $\sigma$ and $A_l$ are respectively the absorption coefficient, the length, the stimulated emission cross-section and the effective beam area of the laser crystal. $\gamma$ is equal to one for an ideal four-level system and $(\sigma_e/\sigma_g)_{sa}$ is the ratio between excited state and ground state absorption cross sections of the saturable absorber. $\chi$ is a factor related to the pump intensity (see reference [13] for details) : the higher the pump intensity, the higher is the $\chi$ factor.

Equation (2) shows that the Q-switching criterion will be more easily fulfilled for high pump intensity as consequently for high brightness laser diodes. Another issue could be to define geometrical parameters properly (length of the media and beam area). However, for a microchip laser, unlike in extended cavities [11, 15, 16] the beam area in the laser is equal to the one in the saturable absorber. Moreover, the lengths of these media are set by other requirements such as pulse duration or repetition rate. It remains a ratio $\xi = \alpha_{sa}\sigma_{sa} / \alpha_l\sigma_l$ related to spectroscopic parameters and doping levels which has to be as high as possible in order to fulfil the Q-switch criterion. The table 1 gives different values of ξ for Nd:YAG, Nd:YVO4 and Nd:GdVO4 (a-cut or c-cut). It shows clearly that Q-switch criterion is more difficult to achieve in Nd:doped vanadate crystals that in Nd:YAG. However, the choice of a c-cut vanadate crystal represents a good compromise between high repetition rate (requiring high emission cross section) and possibility to achieve the passively Q-switched operation.

Up to now, passively Q-switched Nd:doped vanadate lasers have been demonstrated in extended cavities [11, 13-16]. The only passively Q-switched $Cr^{4+}$:YAG/Nd:YVO$_4$ microlaser has been reported by Jaspan et al. in 2000 [17]. The repetition rate was 50 kHz with a pulse duration around 2 ns and 50 mW of average power. However, it was not a monolithic laser.

Among the available vanadate crystals, Nd:GdVO$_4$ [18] had attracted great attention since it presents the same qualities as its isomorph Nd:YVO$_4$ along with a higher thermal conductivity. These unique spectroscopic and thermal properties make this crystal very promising with a view of increasing the power of microchip laser – often limited by thermal considerations. Nd:GdVO$_4$ passively Q-switched microchip lasers are consequently an interesting issue to investigate : we present in the following what is to our knowledge the first monolithic passively Q-switched diode pumped $Cr^{4+}$:YAG/Nd:GdVO$_4$ microchip laser.

## 2. Experimental results

Following the previous consideration, we chose a c-cut crystal Nd:GdVO4 as the ξ parameter is closer to the one of Nd:YAG, favouring the passive Q-switching operation.

The Nd:GdVO$_4$ crystal used in our experiment is 1% doped and 1 mm long. The setup used for our experiment is described on fig.1. The $Cr^{4+}$:YAG saturable absorber was 0.4 mm long and its absorption coefficient is 4 cm$^{-1}$. It was optically contacted on the laser crystal. Lateral dimensions of the chips were 1 mm by 1 mm. One end face of the chip was high reflection coated at 1064 nm and high transmission at 808 nm when the other end face (output coupler, Cr$^{4+}$:YAG side) reflection coefficient is 85% at 1064 nm. The chip was mounted in a copper holder with contact on the upper and lower faces of the crystal only. A Peltier element was used to control its temperature. The pump source was a 10W-fibre-coupled laser diode (LIMO) with a core diameter of 100 µm and a numerical aperture of 0.2. The output of the fibre was imaged with a magnification of one in the microlaser, leading to a pump spot waist radius in the chip of 50 µm. The wavelength of the pump diode was adjusted to maximize the absorption in the microchip laser, which was found to be 74 %. We estimated the $\chi$ factor for the Nd:GdVO$_4$ microlaser to be larger than unity [16] (around 40 at threshold) : the Q-switching criterion was then verified.

In a view of comparison, we used the same intense pumping setup to pump a Cr:YAG/Nd:YAG microlaser with comparable characteristics. The chip was made of 1 mm long Nd:YAG crystal (1.3% doped) optically contacted to a 250 µm long Cr:YAG crystal (with an absorption coefficient of 6 cm$^{-1}$). The measured absorption of this microlaser was 56%. The coated output coupler had a transmission of 85 %.

The figure 2 gives the pulse repetition frequency versus the pump power. Experimentally, all the datas reported here were obtained for stable operation with only one longitudinal mode. It is worth noting that at a pump power of 2W, the repetition rate obtained with our Nd:GdVO$_4$ microchip laser overcame by a factor of two the repetition rate of the Nd:YAG microlaser.

At higher pump power level, one can notice saturation on the experimental repetition rate. In fact, above this level the operation of the microchip became less stable, with increasing temporal jitter. We attribute this phenomenon to gain effect: under strong pumping, the laser oscillated classically on several longitudinal modes. We then got single mode operation back by decreasing the available gain through defocusing, but in this case the repetition rate remained at the same level. Single mode operation cannot be achieved for incident pump powers above 3.1 W with the Nd:GdVO$_4$ chip. The maximum repetition rate we have obtained with this laser is 85 kHz, which is significantly higher that what can be obtained with Nd:YAG in stable operation (from 50 kHz [9] for the best published results to 60 kHz in the present study). The timing jitter was generally measured to be around 1% of the period. The pulse duration measured using a fast photodiode and a 2.5 GHz bandwidth oscilloscope (Tektronix TDS 7254) was around 1.1 ns for Nd:GdVO$_4$ (fig 3a) and 850 ps for Nd:YAG, with no major variation with increasing pump power. For all measurements the output beam is diffraction limited with a M$^2$ factor around 1.15 (see figure 3b). The output beam is nearly linearly polarized for both Nd:GdVO$_4$ and Nd:YAG microlasers, with a ratio 1:9 between two orthogonal states of polarization. This can be attributed to the influence of thermal gradients in the crystals due to the asymmetric cooling used in the experiment.

The average output powers are given on the figure 4. For the Nd:GdVO$_4$ microlaser, we obtained around 400 mW in longitudinal single mode operation for 3.1 W of incident pump power (2.3 W of absorbed pump power), corresponding to pulse energy of 4.2 µJ and a peak power of nearly 4 kW. The optical conversion efficiency was around 18 % with a slope efficiency of 20 %. The Cr:YAG/Nd:YAG microchip laser produced the same amount of average power but with a slightly lower slope efficiency.

In conclusion, we present in this paper what is to our knowledge the first monolithic Nd:GdVO$_4$ microchip laser passively Q-switched with a Cr:YAG saturable absorber. We showed that Q-switching operation is possible with Nd:GdVO$_4$ with adequate pumping conditions, that is to say under high intensity pumping. We obtained high repetition rate operation around 80 kHz with 1.1 ns long pulses in transverse and single longitudinal mode operation, which represents an significant improvement compared to Nd:YAG microchip lasers. Moreover, this kind of vanadate microlaser opens the way to interesting improvements for the already existing Nd:YAG/Nd:YVO4 MOPA systems (using a Nd:YAG passively Q-switched microlaser) because it allows a perfect overlap between the wavelength of the oscillator and the peak emission cross section of the high gain vanadate amplifier. The very simple, robust, low cost and compact design makes this kind of microchip laser very promising for a large number of applications where high repetition rates are required.

1. <u>Figures captions</u>

Table 1 : comparison between Nd:YAG, Nd:YVO$_4$ and Nd:GdVO$_4$.

Fig.1 : experimental setup.

Fig.2 : repetition rate versus incident pump power. The theoretical curves are obtained with the formula (1).

Fig 3: (a) typical pulse shape for the Nd:GdVO$_4$ microchip laser. (b) spatial beam profile and typical M² measurement.

Fig 4 : average power versus incident pump power

Table 1 : comparison between Nd:YAG, Nd:YVO$_4$ and Nd:GdVO$_4$.

|  | Nd:YAG | Nd:YVO$_4$ a-cut | Nd:YVO$_4$ c-cut | Nd:GdVO$_4$ a-cut | Nd:GdVO$_4$ c-cut |
|---|---|---|---|---|---|
| Absorption at peak 808 nm (cm$^{-1}$) | 7.1 | 51 | 10 | 45 | 11 |
| Emission cross section at peak 1064 nm (cm²) | 2.1 10$^{-19}$ | 12.5 10$^{-19}$ | 6 10$^{-19}$ | 12.5 10$^{-19}$ | 6.1 10$^{-19}$ |
| ξ parameter[a] | 11.5 | 0.28 | 2.87 | 0.31 | 2.56 |

[a]The ξ parameter is defined in text (we take for Cr:YAG $\alpha_{sa}$ = 4 cm$^{-1}$ and $\sigma_{sa}$ = 4.3 10$^{-18}$ cm²).

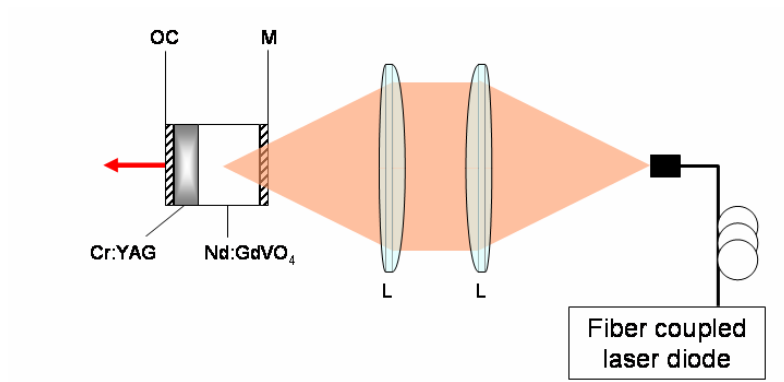

Fig.1 : experimental setup.

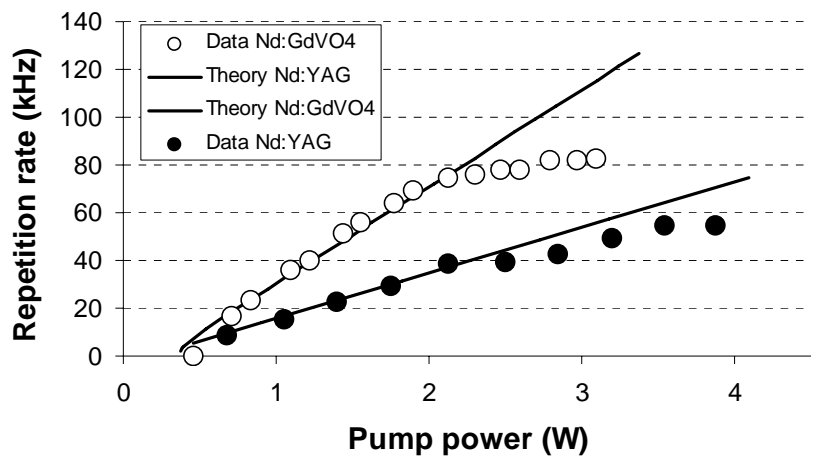

Fig.2 : repetition rate versus incident pump power. The theoretical curves are obtained with the formula (1).

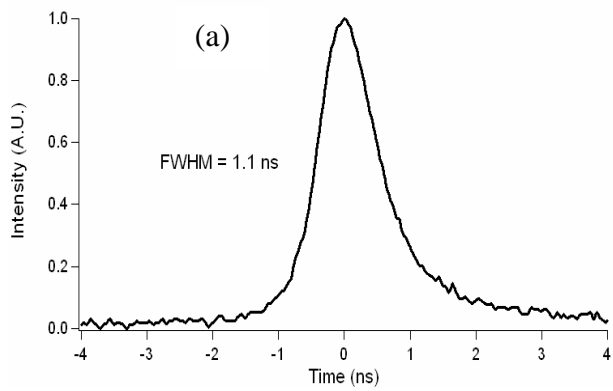 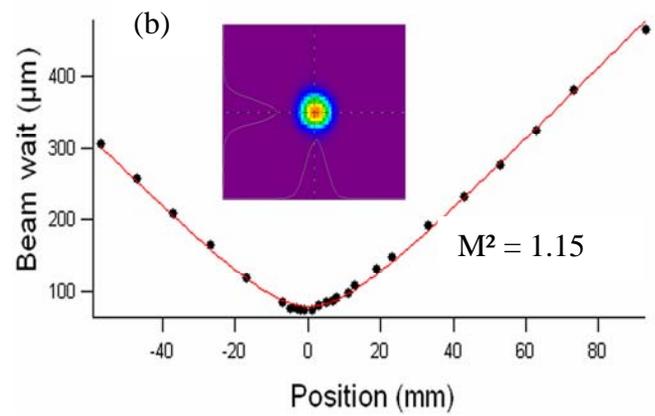

Fig 3: (a) typical pulse shape for the Nd:GdVO$_4$ microchip laser. (b) spatial beam profile and typical M² measurement.

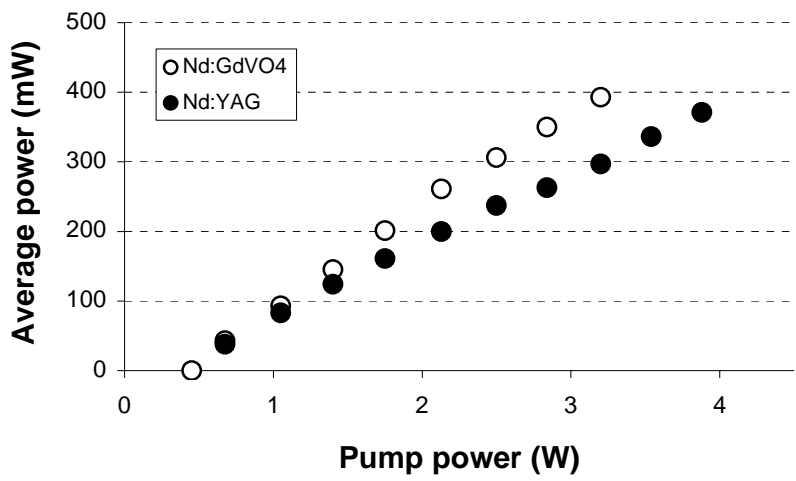

Fig 4 : Average power versus incident pump power